\documentstyle[preprint,aps]{revtex}
\begin{document}
\draft
\preprint{}
\title{Restrictions on Magnetic Charge from Quantized Angular Momentum}
\author{D. Singleton}
\address{Department of Physics, Virginia Commonwealth University, 
Richmond, VA 23284-2000}
\date{\today}
\maketitle
\begin{abstract}
Using the result that an electric charge - magnetic charge system
carries an internal field angular momentum of $e g / 4 \pi$ we
arrive at two restrictions on magnetic monopoles via the requirement of 
angular momentum quantization and/or conservation. First we show
that magnetic charge should scale in the opposite way from
electric charge. Second we show that {\it free}, {\it unconfined} 
monopoles seem to be inconsistent with quantized angular momentum 
when one considers a magnetic charge in the vicinity of more than
one electric charge.
\end{abstract}
\newpage
\narrowtext

\section{Running of the Magnetic Coupling} 

One of the most unusal results of including magnetic charges
in electromagnetism is that an isolated electric charge - 
magnetic charge system carries an angular momentum in the
electric and magnetic fields of the system. If the electric charge
has a magnitude $e$ and the magnetic charge a magnitude $g$ the
angular momentum carried in the ${\bf E}$ and ${\bf B}$ fields
is
\begin{equation}
\label{angmom}
{\bf L} _{fields} = {e g \over 4 \pi} {\hat {\bf r}}
\end{equation}
where ${\hat {\bf r}}$ is the unit vector pointing from the
electric charge to the magnetic charge \cite{jackson}.
This result, along with the quantum mechanical requirement
that all angular momentum come in integer units of $\hbar / 2$
is the best formulation independent way of arriving at the 
Dirac quantization condition that $e g / 4 \pi = n (\hbar / 2)$
$n = 1, 2, 3, ...$ \cite{saha}. (It is formulation independent in that 
it only requires the assumption of the Coulombic form of the ${\bf E}$
and ${\bf B}$ fields surrounding the electric and
magnetic charges. It is not necessary to assume either the
string approach of Dirac \cite{dirac} or the fiber-bundle
approach of Wu and Yang \cite{wu}). Using Eq. (\ref{angmom})
and the conservation and quantization of angular momentum it is 
possible to determine how the magnetic coupling $g$ scales with energy.

The running with energy of the coupling of a theory is determined
by the beta function which is defined as
\begin{equation}
\label{beta}
\beta _{\lambda} = \mu {\partial \lambda \over \partial \mu}
\end{equation}
where $\lambda$ is the coupling and $\mu$ is an arbitrary energy
scale. The electric coupling $e$, and the magnetic coupling $g$,
as with any other coupling, should have a dependence 
on the energy scale $\mu$. The usual way of finding 
$e (\mu)$, for example, is to calculate the beta function
perturbatively to the desired order in perturbation theory, plug
this into the left hand side of Eq. (\ref{beta}), 
and then solve the resulting differential
equation for $e (\mu)$. Since the magnetic coupling $g(\mu)$ is
nonperturbatively large (from the Dirac quantization condition 
and $e$ being perturbatively small) it is difficult to calculate 
the magnetic beta function, $\beta _g$, in the usual way. 
For an isolated electric charge - magnetic charge
system the quantity $e (\mu) g(\mu) / 4 \pi$ must be a constant
since otherwise angular momentum conservation would be violated.
Additionally $e(\mu)$ and $g(\mu)$ must scale in a related way so
that the field angular momentum is always equal to some integer
multiple of $\hbar / 2$. Since $e (\mu)$ and $g (\mu)$ run in
a continuous way the only way to satisfy this requirement of
quantized angular momentum is for $e (\mu) g(\mu) / 4 \pi$ to
remain equal to whatever integer multiple of $\hbar / 2$ it
was equal to at the reference energy scale $\mu _0$. Thus
taking the Dirac condition $e(\mu) g(\mu) / 4 \pi = n (\hbar / 2)$ 
differentiating both sides by $\mu$ and multiplying by 
$4 \pi \mu$ one gets
\begin{equation}
e  \left( \mu {\partial g \over \partial \mu} \right) 
+ g \left( \mu {\partial e \over \partial \mu} \right) = 0
\end{equation}
which with the help of Eq. (\ref{beta}) can be written as
\begin{equation}
\label{betarel}
\beta _g = - {g \over e} \beta _e
\end{equation}
The subscripts indicate the magnetic or electric coupling.
The minus sign in Eq. (\ref{betarel}) shows that the magnetic
coupling and electric coupling should run in opposite ways. Moreover
Eq. (\ref{betarel}) should be valid to any order of perturbation
theory since it just depends on the requirement that angular 
momentum be conserved and quantized. In order to make any
detailed statements about the scaling of $e(\mu)$ or $g(\mu)$ one must
be able to calculate either $\beta _g$ or $\beta _e$. Taking as an 
example spinor QED where $\beta _e = e^3 / 12 \pi ^2$, and using the Dirac
quantization condition to replace $e$ with $g$ via $e = 2 \pi / g$
(where we have specialized to the $n=1$ case and set $\hbar =1$) 
it is found that Eq. (\ref{betarel}) becomes $\beta _g = 
{-1 \over 3 g}$. Solving this for $g(\mu)$ yields
\begin{equation}
\label{grun}
g^2 (\mu) = g^2 (\mu _0) - {2 \over 3} \ln \left( {\mu \over \mu_0}
\right)
\end{equation}
where $\mu _0$ is the reference energy scale. From Eq. (\ref{grun})
it is easy to see that the magnetic coupling decreases with energy
as opposed to the electric coupling, $e (\mu)$, which increases
with energy. The decrease is logarithmically slow so that
the magnetic coupling will still remain enormous out to energy
scales beyond which the perturbative calculation of $\beta _e$ 
can be trusted, and far out of reach of any current or planned 
accelerators. This example, or a more realistic extension using
$\beta _e$ of the Standard Model, depend on the implicit assumption
that monopoles can be ignored in the perturbative calculation
of $\beta _e$. Recently it has been shown that experimentally
measured Standard Model parameters do not show any sign of the
effect of virtual monopoles up to roughly 1 TeV \cite{der}. The
harder  and more interesting question of how to handle the
situation when virtual monopoles do start to effect $\beta _e$
is left unanswered by the above results, but it is still somewhat
interesting that the scaling of the magnetic coupling can be 
determined in a region where its strength is nonperturbatively large.

Several other authors \cite{deans} \cite{jengo} have investigated
quantum field theories with magnetic charge, and have 
discussed the renormalization of the Dirac quantization condition
as well as the running of the magnetic coupling. In Ref. \cite{deans}
it is claimed that magnetic and electric couplings should 
scale in the same way, while Ref. \cite{jengo} comes to the different
conclusion that they should scale in
the opposite way. Our results agree with those of Ref. \cite{jengo},
but our argument is different. Rather than using a perturbative, 
Feynman diagrammatic approach we arrive at our results based on
angular momentum conservation and quantization. Thus,
although our result is the same as that of Ref. \cite{jengo}, our 
arguments should remain valid to all orders of perturbation theory. 

\section{Monopole and Two Electric Charges}

From Eq. (\ref{angmom}) one sees that an electric charge - magnetic
charge system carries an internal angular momentum due to the
${\bf E}$ and ${\bf B}$ fields of the particles, which has a magnitude
$e g / 4 \pi$ and points from the electric charge toward the magnetic
charge. Now we will consider what happens
when the monopole is in the presence of two electric charges. For 
definiteness we place a positive electric charge, $+e$, at the
origin, a negative electric charge, $-e$, along the z-axis a distance
$d > 1$ away from the origin, and we place the magnetic charge, $g$,
at an arbitrary point on the unit circle in the $xz$ plane so
that its position is given by the polar angle $\theta$. The angular 
momentum of the $+e g$ part of this system points from $+e$ to $g$ and is
\begin{equation}
\label{l1}
{\bf L} _1 = {e g \over 4 \pi} \left( \cos \theta {\hat {\bf x}}
+ \sin \theta {\hat {\bf z}} \right)
\end{equation}
The angular momentum of the $- e g$ part of the system points from
$g$ to $-e$ and is
\begin{equation}
\label{l2}
{\bf L}_2 = {e g \over 4 \pi} \left( {- \cos \theta \over
\sqrt{d^2 +1 -2d \sin \theta}} {\hat {\bf x}} + {d -\sin \theta
\over \sqrt{d^2 +1 -2d \sin \theta}} {\hat {\bf z}} \right)
\end{equation}
The total angular momentum of this system of two electric charges
and one magnetic charge is then simply the sum of Eqs. (\ref{l1})
and (\ref{l2})
\begin{equation}
\label{lt}
{\bf L}_{tot} = {\bf L}_1 + {\bf L}_2
\end{equation}
Now the magnitude of ${\bf L}_{tot}$ varies continuously from
$2 (e g / 4 \pi)$ at $\theta = \pi / 2$, to $0$ at $\theta = 3 \pi
/2$. Thus, classically, any value for $\vert {\bf L}_{tot} \vert$
between $e g / 2 \pi$ and $0$ is possible depending on the choice of
$\theta$. If one requires that a particular value of $\vert 
{\bf L}_{tot} \vert$, for a particular $\theta$, equal some integer 
multiple of $\hbar / 2$ then all the continuously connected values
of $\vert {\bf L}_{tot} \vert$ at neighboring values of $\theta$
will not equal an integer multiple of $\hbar / 2$, in violation of
the requirement that angular momentum be quantized. Further for
$\theta = 0$ or $\pi$ one finds that
\begin{equation}
\vert {\bf L}_{tot} \vert = {e g \over 4 \pi} \sqrt{2 - {2 \over
\sqrt{d^2 -1}}}
\end{equation}
so that $\vert {\bf L}_{tot} \vert$ also has some dependence on
$d$ as well as on $\theta$. The only way for the total angular
momentum of this system to be quantized would be for the initial
positions of the particles (as determined by $\theta$ and $d$)
to take only certain discrete values ({\it i.e.} for the positions
to be quantized). Since there is no apparent mechanism that requires
any such positional ``quantization'' of the initial configuration of
our system we conclude that such a configuration (or more precisely 
the free monopole in the configuration) seems to be inconsistent with 
the quantization of angular momentum. One may worry about the use of the
classical Coulomb fields for the electric and magnetic charges in
the above development (especially in deriving Eq. (\ref{angmom})). 
However, the field angular momentum given in Eq. (\ref{angmom})
is independent of the distance between the charges so that we may
take all the charges to be arbitrarily far away from one another
({\it e.g.} have the magnetic charge placed on a circle of radius
$R$ instead of unit radius such that $d \gg R \gg 1$) without
affecting the result of Eq. (\ref{angmom}). Under these
conditions one is certainly justified in using the Coulomb form
for the electric and magnetic fields of the particles, and ignoring
any quantum corrections.

Since electric charges are known to exist, the 
above argument seems to imply that free magnetic charges
do not exist. Despite this it may still be possible
that monopoles exist, if they are permanently confined
much in the same way that color charged quarks are postulated to
be permanently confined. This seems reasonable in light of the
fact that, from the Dirac condition, the monopole coupling is
an order of magnitude, or more, larger than the equivalent QCD
coupling which is thought to confine quarks. In addition 
Wilson's \cite{wilson} original lattice gauge theory paper
on confinement uses a U(1) gauge field to argue for confining 
behaviour, so that it is not the Abelian (QED) versus 
non-Abelian (QCD) character of an interaction
that is of chief importance but the strength of the
coupling. It may be argued that Wilson's results can not be
applied to magnetic charge since it is the electric charge 
which is the gauge charge and is directly coupled to the
U(1) gauge boson. However, as pointed out in Ref. \cite{baker},
it is just as easy to make the magnetic charge the gauge
charge, which is directly coupled to the U(1) gauge boson, by
introducing a four vector potential, $C_{\mu} = (\phi _m ,
{\bf C})$, which is dual to the usual four vector potential
$A_{\mu}$. Then by defining a new field strength tensor and
its dual as
\begin{equation}
\label{fstdual}
G_{\mu \nu} = \partial _{\mu} C_{\nu} - \partial _{\nu} C_{\mu}
\; \; \; \; \;
{\cal G}_{\mu \nu}  = {1 \over 2} \epsilon _{\mu \nu \alpha \beta}
G^{\alpha \beta}
\end{equation}
one can obtain the ${\bf E}$ and ${\bf B}$ fields as
\begin{equation}
E_i = - {\cal G}^{i0} \; \; \; \; \; B_i = G^{i0}
\end{equation}
(In three vector notation this becomes ${\bf E} = - \nabla \times
{\bf C}$ and ${\bf B} = - \nabla \phi _m - (1/c) (\partial {\bf C}
/ \partial t)$). In this dual formulation it is the magnetic
charge with is the gauge charge and the electric charge which finds
itself on the end of a Dirac string. Thus Wilson's arguments for
confinment should apply equally well to magnetic charge in this
dual formulation of electrodynamics. Although in principle either
formulation of electrodynamics (in terms of $A_{\mu}$ or $C_{\mu}$)
is equally satisfactory, in practice it much easier
to deal with whatever charge is directly coupled to the U(1) gauge
boson. Since electric charges are known to exist while
magnetic charges have never been seen experimentally it is more
practical to do electrodynamics with electric charge directly coupled
to the U(1) gauge boson. (An interesting possibility in electromagnetism
with both electric and magnetic charges is to use both $A_{\mu}$ and
$C_{\mu}$, and have one of the extra ``photons'' ({\it i.e.} $A_{\mu}$
or $C_{\mu}$) hidden through some form of symmetry breaking \cite{sing}).

\section{Conclusions}

Using the result that a magnetic charge placed in the field
of one or more electric charges produces a field
angular momentum, and that angular
momentum must be conserved and quantized, we have given two
restrictions on the behaviour of monopoles. First Eq. (\ref{betarel})
shows that the magnetic coupling will scale in the opposite way
from the electric coupling. Although the running of the
magnetic coupling, as given in the example of spinor QED,
is logarithmically slow, so that the magnetic coupling
will still be enourmous out to any energy scale that may be
reasonably reached at any future accelerator, it is interesting
that one can say anything about the scaling of a coupling that is
nonperturbatively large. This result is not new \cite{jengo},
but we have arrived at it using only angular momentum conservation
and quantization so that our result, as given in Eq. (\ref{betarel}),
should be good to any order. Second, by considering the field
angular momentum of a system of two electric charges and one 
monopole we have found that it is not possible to have the
field angular momentum quantized to some integer multiple
of $\hbar /2$ for every possible initial position of
this system ({\it i.e.} angular momentum can not be quantized
since the initial positions of the charges need not be
``quantized''). Since the quantization of angular momentum
is a fundamental requirement of quantum mechanics this appears
to imply that free, unconfined monopoles are incompatiable with quantum 
mechanics and therefore do not exist. A possible evasion of this 
argument is that even if there are no free monopoles they may exist 
and be permanently confined, in the same way quarks are thought to
be permanently confined. (This confinement hypothesis for
magnetic charge also fits in with the scaling of the
magnetic coupling discussed in the first section). The dual 
formulation of QED \cite{baker}, where magnetic charge is the
gauge charge, combined with the lattice gauge theory confinement arguments
of Wilson \cite{wilson} also help to give strength to this confinement 
hypothesis of magnetic charge. This would also give an explanation
as to why free monopoles have evaded experimental detection, just as
free quarks have evaded detection.

Finally one may ask how the above arguments apply to 't Hooft-Polyakov
\cite{thooft} monopoles. It would appear that since our arguments
depend on the electric charge - magnetic charge system carrying
a field angular momentum that they should apply to these topological
magnetic charges as well. It has been shown that the composite 
system of a 't Hooft-Polyakov monopole and a non-Abelian
test particle also carries a field angular momentum
\cite{rebbi}. However in this case things may be a bit more
delicate since it is not clear that one can use the principle
of superposition as freely as in the Abelian case.
For example, if the Abelian monopole were replaced by a 't Hooft-Polyakov
monopole, and the two electric charges by non-Abelian charges, and
if the superposition were valid, then a conflict with quantized
angular momentum would also arise in this non-Abelian system. However
non-Abelian gauge theories are non-linear so the superposition
principle is not valid. Without further analysis the best that
can be done is to speculate that since the
field angular momentum is again independent of the distance between
the charges that all the charges could again be separated by
arbitrarily large distances so that superposition 
would be approximately valid.

\section{Acknowledgements} The author would like to thank Justin
O'Neill and Miwa Rucker for help and encouragement during the
writing of this paper.

\end{document}